\documentclass[aps,prl,twocolumn,amsmath,amssymb,reprint,longbibliography,superscriptaddress]{revtex4-1}

\usepackage[utf8]{inputenc}
\usepackage{amsmath}
\usepackage{graphicx}
\usepackage{dcolumn}
\usepackage{bm}
\usepackage{makeidx}
\usepackage[alsoload=synchem]{siunitx} 
\usepackage{physics}
\usepackage{hyperref}
\usepackage{cleveref}	
\usepackage[usenames, dvipsnames]{color}

\begin{document}
\title{Observation of 2$e$-periodic Supercurrents in Nanowire Single-Cooper-Pair Transistors}
\author{Jasper van Veen}
\author{Alex Proutski}
\affiliation{QuTech, Delft University of
Technology, 2600 GA Delft, The Netherlands}
\affiliation{Kavli Institute of Nanoscience, Delft University of
Technology, 2600 GA Delft, The Netherlands}

\author{Torsten Karzig}
\author{Dmitry I. Pikulin}
\author{Roman M. Lutchyn}
\affiliation{Station Q, Microsoft Corporation, Santa Barbara, California
93106-6105, USA}

\author{Jesper Nyg{\aa}rd}
\affiliation{Center for Quantum Devices, Niels Bohr Institute,
University of Copenhagen, 2100 Copenhagen, Denmark}

\author{Peter Krogstrup}
\affiliation{Center for Quantum Devices and Microsoft Quantum Lab Copenhagen, Niels Bohr Institute,
University of Copenhagen, 2100 Copenhagen, Denmark}

\author{Attila Geresdi}
\affiliation{QuTech, Delft University of
Technology, 2600 GA Delft, The Netherlands}
\affiliation{Kavli Institute of Nanoscience, Delft University of
Technology, 2600 GA Delft, The Netherlands}

\author{Leo P.~Kouwenhoven}
\affiliation{Microsoft Station Q Delft, Delft University of Technology, 2600 GA Delft, The
Netherlands}
\affiliation{QuTech, Delft University of
Technology, 2600 GA Delft, The Netherlands}
\affiliation{Kavli Institute of Nanoscience, Delft University of
Technology, 2600 GA Delft, The Netherlands}

\author{John D.~Watson}
\email{To whom correspondence should be addressed;
E-mail:  john.watson@microsoft.com}
\affiliation{Microsoft Station Q Delft, Delft University of Technology, 2600 GA Delft, The
Netherlands}

\date{\today}
\maketitle

\textbf{Parity control of superconducting islands hosting Majorana zero modes (MZMs) is required to operate topological qubits made from proximitized semiconductor nanowires. We, therefore, study parity effects in hybrid InAs-Al single-Cooper-pair transistors (SCPTs) as a first step. In particular, we investigate the gate-charge supercurrent modulation and observe a consistent 2$e$-periodic pattern indicating a general lack of low-energy subgap states in these nanowires at zero magnetic field. In a parallel magnetic field, an even-odd pattern develops with a gate-charge spacing that oscillates as a function of field demonstrating that the modulation pattern is sensitive to the presence of a single subgap state. In addition, we find that the parity lifetime of the SCPT decreases exponentially with magnetic field as the subgap state approaches zero energy.  Our work highlights the important role that intentional quasiparticle traps and superconducting gap engineering would play in topological qubits that require quenching of the island charge dispersion.} 

\begin{figure}[t]
\begin{center}
\includegraphics[width=0.5\textwidth]{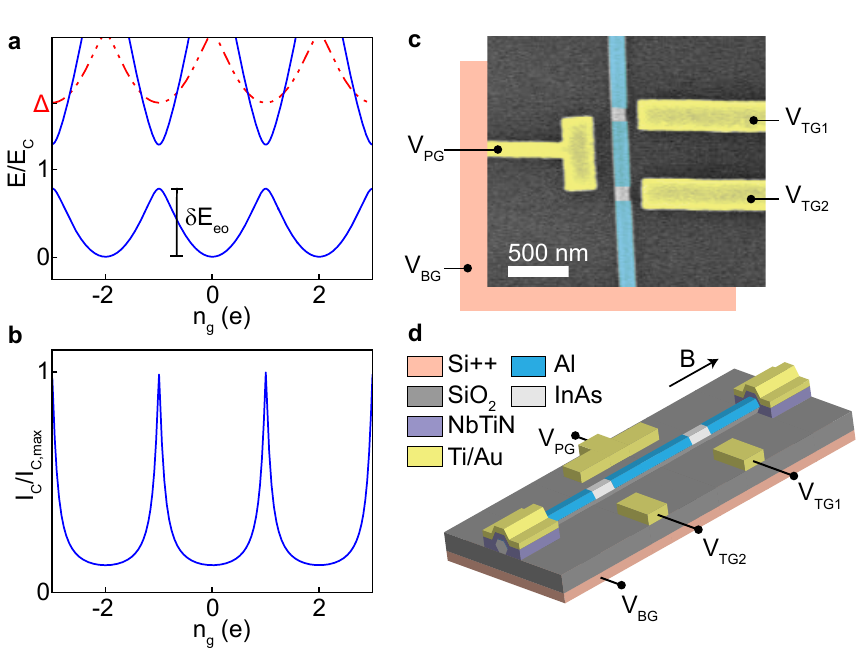}
\caption{\textbf{Theoretical background and device layout.} \textbf{(a)}  The band structure of a SCPT as a function of gate-charge $n_{g}$ for $\phi=0$, $\Delta/E_C = 1.5$, and $E_{J}/E_{C} = 0.25$. The charge dispersion of the odd parity branch (in red) is displaced from the even parity branch (in blue) by the superconducting gap $\Delta$ and $\Delta n_g = 1$. The amplitude of the ground state charge dispersion is denoted by $\delta E_\textbf{eo}$ \textbf{(b)} The corresponding critical current modulation as a function of $n_{g}$. \textbf{(c)} A false-coloured scanning electron micrograph of a nanowire SCPT. The etched regions in the Al shell define the junctions and the island. By applying voltages to the electrostatic gates, we can tune the electrochemical potential (with $V_{BG}$), the junction transparency (with $V_{TG1,2}$) and the charge occupation of the island (with $V_{PG}$). \textbf{(d)} 3-dimensional device schematic. The nanowire is deterministically placed on top of a SiO$_{2}$/Si++ substrate. It is then contacted by a stack of NbTiN and Ti/Au \SI{1}{\micro\meter} away from the etched regions. Finally, the local gates are deposited. The arrow indicates the direction of the magnetic field for the data presented in Fig. \ref{Fig4}.}
\label{Fig1}
\end{center}
\end{figure}

The interplay of charging energy $E_C$ and the superconducting gap $\Delta$ leads to the surprising result that the electrical transport in a mesoscopic superconducting island containing a macroscopic number of electrons is sensitive to the addition or removal of a single electron \cite{matveev1993parity, averin1992single, grabert2013single, averin1986coulomb}. This parity effect has been extensively studied in Al-AlO$\mathrm{_x}$ SCPTs by measurements of the 2$e$-periodic gate-charge modulation of the Coulomb peak spacing, the ground state charge, and the switching current \cite{Mooij19902e, joyez1994parity, tuominen1992parity, lafarge1993evenodd, eiles1994supercurrent,mannik2004effect,naaman2006time,ferguson2006microsecond}. In recent experiments, the presence of MZMs in hybrid semiconductor-superconductor nanowires was inferred from the field-induced 1$e$ Coulomb blockade periodicity, illustrating the utility of this periodicity in understanding the low-energy spectrum of mesoscopic superconducting islands \cite{higginbotham2015parity, albrecht2016exponential, sherman2017normal, albrecht2017transport, shen2018parity, o2018hybridization, Fu2010, vanHeck16}. In contrast with these previous studies which utilized devices with normal metal leads, we investigate parity effects in  gate-tuneable nanowire SCPTs which have superconducting leads. We note that the device in Ref. \cite{o2018hybridization} also has superconducting leads but was mainly operated in a large in-plane field in which the leads were effectively normal. We study the tunnel gate, temperature, and parallel magnetic field dependence of the switching current modulation. These experiments not only give new insights into quasiparticle dynamics but also represent a first step towards implementing recent Majorana-based qubit proposals which require Josephson coupling to the leads to enable parity-to-charge conversion for MZM manipulation and readout \cite{van2012coulomb, hyart2013flux, aasen2016milestones, hell2016time, litinski2017combining}. 

The Hamiltonian of a SCPT consists of three terms: $H = H_C+H_J+H_\text{BCS}$. The Coulomb term, $H_C = E_C(n-n_g)^2$, stabilizes the excess charge $n$ on the island which can be changed by varying the gate-charge $n_g$. The effective charging energy $E_C = e^2/2C$ is given in terms of the electron charge $e$ and a generalized capacitance $C$ that resembles the island's geometric capacitance in the limit of a weakly coupled island \cite{Ambegaokar82,Larkin83,Eckern84,Lutchyn07}. The Josephson term for symmetric junctions $H_J = E_J\cos\left(\phi/2 \right) \sum_n |n\rangle\langle n+2| + h.c.$, with $E_J$ the Josephson energy and $\phi$ the superconducting phase difference across the island, couples adjacent, equal-parity states and results in energy level anti-crossings when states with the same parity are degenerate. The third term describes the spectrum of the gapped BCS quasiparticles thus resulting in an energy offset $\Delta$ for the odd ground state due to an unpaired electron in the superconductor. Figure \ref{Fig1}a shows the resulting band structure of a SCPT. The corresponding gate-charge modulation of the critical current is shown in Fig. \ref{Fig1}b. We denote the amplitude of the (even) ground state charge dispersion $E_\text{gs}(n_g)$ with $\delta E_\text{eo}=E_{\text{gs}}(n_{g}=1)-E_{\text{gs}}(n_{g}=0)$. When $\Delta >  \delta E_\text{eo}$ the ground state is always even. Consequently, the switching current modulation will be 2$e$-periodic at $T = 0$ in this simple model.

Quasiparticle poisoning, however, affects this 2$e$-periodic modulation. Previous studies have illustrated three important timescales, namely the poisoning rate $\Gamma_\text{in}$ at which quasiparticles in the lead tunnel to the island, the non-equilibrium unpoisoning rate $\Gamma^{\text{neq}}_\text{out}$ at which non-equilibrium quasiparticles on the island tunnel out to the leads, and the relaxation rate $1/\tau$ at which non-equilibrium quasiparticles on the island relax to the gap edge or subgap states \cite{aumentado2004nonequilibrium,lutchyn2007kinetics, shaw2008kinetics,court2008quantitative}. While the relaxation is important for the quasiparticle dynamics, the thermodynamics of the system can be described by equilibrium poisoning and unpoisoning rates $\Gamma_\text{in}$ and $\Gamma_\text{out}$ alone; therefore, we leave the implications of relaxation in our devices to the discussion section below. The ratio $\Gamma_\text{in}/\Gamma_\text{out}$ gives the relative occupation between the even and odd parity states in equilibrium $p_\text{odd}/p_\text{even}$. If $\Gamma_\text{in}/\Gamma_\text{out} \approx 1$ as is expected to occur at high temperature, the switching current modulation deviates from 2$e$ periodicity and exhibits a 1$e$ periodicity instead.

Figure \ref{Fig1}c presents a scanning electron micrograph (SEM) of one of our SCPTs, and a 3-dimensional schematic of the device is shown in Fig. \ref{Fig1}d. The SCPTs are fabricated from InAs nanowires covered with a thin aluminium shell on two of their facets. It has been shown that this material combination results in a hard, induced superconducting gap in the nanowire \cite{krogstrup2015epitaxy, chang2015hard}. The aluminium shell is etched in two regions along the nanowire in order to define the island together with the two Al-InAs-Al Josephson junctions. The wire is contacted \SI{1}{\micro\meter} away from each junction by NbTiN/Ti/Au contacts which are expected to act as quasiparticle traps due to the presence of normal metal and the large sub-gap density of states in NbTiN \cite{van2015one}. Voltages $V_{TG1}$ and $V_{TG2}$ applied to the side gates tune the transparency of the weak links while the plunger gate voltage $V_{PG}$ tunes the chemical potential of the island, and the global backgate voltage $V_{BG}$ tunes the chemical potential of the whole the system. The SCPTs are glued to the cold finger of a dilution refrigerator with a base temperature of 27 mK. We report on six devices in total; in the main text we present data on a device with a 500 nm long island (see Table S1 of the Supplementary Information for an overview of all the devices). Unless otherwise indicated, the presented data were obtained at 27 mK and at zero magnetic field. 

\begin{figure}[t]
\includegraphics[width=0.5\textwidth]{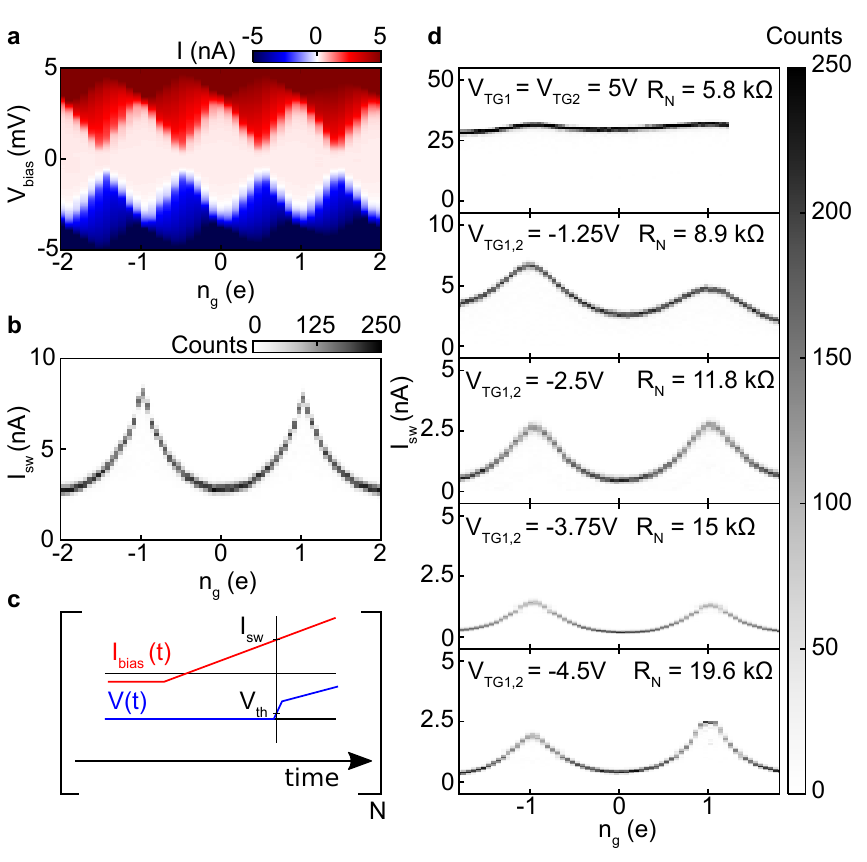}
\caption{\textbf{Gate dependence of the 2$e$-periodic switching current modulation.} \textbf{(a)} Charge stability diagram measured in the strongly Coulomb-blockaded regime with $\Delta = \SI{180}{\micro\electronvolt}$ and $E_C^{0} = \SI{1.5}{\milli\electronvolt}$. \textbf{(b)} Histogram of the $2e$-periodic switching current $I_{sw}$ in the weakly Coulomb-blockaded regime, indicating that $\Delta > \delta E_\text{eo}$. At this gate setting, $V_{TG1}$ = -4.1 V and $V_{TG2}$ = -5.7 V, $R_N = \SI{14.8}{\kilo\ohm}$. \textbf{(c)} Schematic representation of the sawtooth current bias waveform (in red) used to obtain the $I_{sw}$ histograms and the resulting voltage across the SCPT (in blue). The switching current $I_{sw}$ is recorded when the voltage drop on the SCPT reaches a threshold value $V_{th}$. \textbf{(d)} Switching current histograms for varying normal state resistance. The normal state resistance is calculated as the average over the $n_g$ range at high bias. Note the change of vertical scale for the two topmost panels.  The peak height asymmetry seen for $R_N =  \SI{8.9}{\kilo\ohm}$ and  $R_N = \SI{19.6}{\kilo\ohm}$ is due to cross coupling between the tunnel barriers and $V_{PG}$.}
\label{Fig2}
\end{figure}

\begin{figure}[ht]
\includegraphics[width=0.5\textwidth]{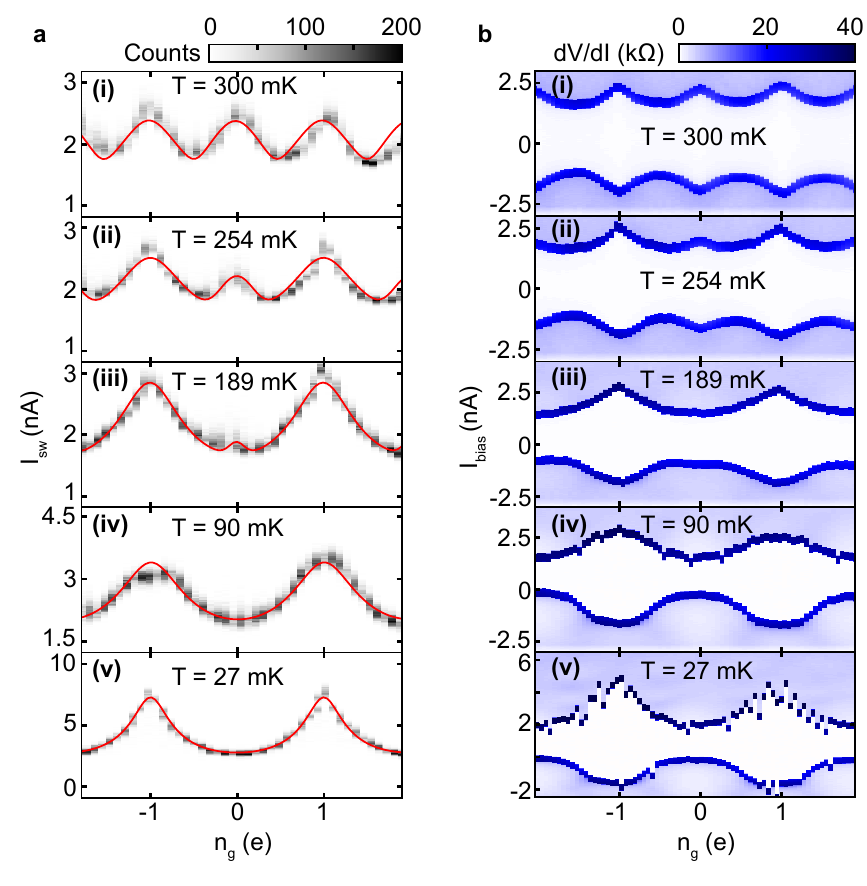}
\caption{\textbf{Temperature dependence.} \textbf{(a)} Switching current modulation as a function of temperature. The experimental histograms shown in grayscale are overlaid by the theoretical fit to the average switching current $\left\langle I_\text{sw}\right\rangle$ (red curves). Individual fits are for different values of $\Delta$, $E_J$, and $E_C$. The resulting values for the parameters are $\Delta \approx \SI{220}{\micro\electronvolt}$, $E_J\approx \SI{43}{\micro\electronvolt}$, and $E_C \approx \SI{160}{\micro\electronvolt}$. \textbf{(b)} $dV/dI$ data for the same temperature range as in \textbf{(a)} obtained from numerical derivation of the I-V curves. At elevated temperatures the overdamped $dV/dI$ data shows a similar behavior as the histograms in \textbf{(a)} with local maxima appearing at even $n_g$ at $T \approx 189$ mK and a fully 1$e$ periodic modulation at $ T^* \approx 300$ mK. At low temperatures the junction is in the underdamped regime as indicated by the asymmetric $dV/dI$ and the increased fluctuations due the absence of self-averaging.}
\label{Fig3}
\end{figure}

\section{Results}
\textbf{Coulomb blockade and switching current histograms.} We first tune the device into Coulomb blockade by increasing the heights of the tunnel barriers separating the island from the leads. The clear, regular Coulomb diamonds shown in Fig. \ref{Fig2}a demonstrate the creation of a single, well-defined island. Moreover, a 1$e$-periodic conductance modulation appears when $e\left|V_b\right| > 4\Delta$  and transport through the island is dominated by quasiparticles which enables us to identify the gate voltage periodicity corresponding to 1$e$ \cite{hadley19983}. The current at lower bias voltages is too small to resolve in the Coulomb blockade regime since it involves Cooper pair transport and is therefore higher order in the tunneling. Finally, we extract a superconducting gap $\Delta = \SI{180}{\micro\electronvolt}$ and the geometric charging energy $E_C^{0} = \SI{1.5}{\milli\electronvolt}$ from the observed diamonds.

In order to generate a measurable supercurrent, we lower the tunnel barriers in order to increase $E_J$ which simultaneously suppresses $\delta E_\text{eo}$. The switching current is recorded by triggering on the voltage step in the I-V curve as illustrated in Fig. \ref{Fig2}c; this is repeated $N$ times for each $n_g$ to gather statistics, typically $N =$ 100 to 500.  Figure \ref{Fig2}b shows the resulting switching current histogram which is 2$e$-periodic, indicating that in this regime the charge dispersion has decreased at least an order of magnitude to the point that $\delta E_\text{eo} < \Delta$, consistent with the observed charging energy renormalization in a nanowire island with normal leads \cite{sherman2017normal}. 

To establish that our observed 2$e$ periodicity is indeed robust, we investigate the gate-charge modulation for a wide range of gate settings, as is shown in Fig. \ref{Fig2}d.  We characterize each gate setting by the normal state resistance of the device. Figure \ref{Fig2}d shows that the modulation is observed for $R_N$ ranging from 5.8 to $\SI{19.6}{\kilo\ohm}$. At $R_N = \SI{5.8}{\kilo\ohm}$, the switching current was only modulated by 5\%, indicating that the device is in the Josephson dominated regime where $E_J > \delta E_\text{eo}$.   

\begin{figure*}[ht]
\begin{center}
\includegraphics[width=1\textwidth]{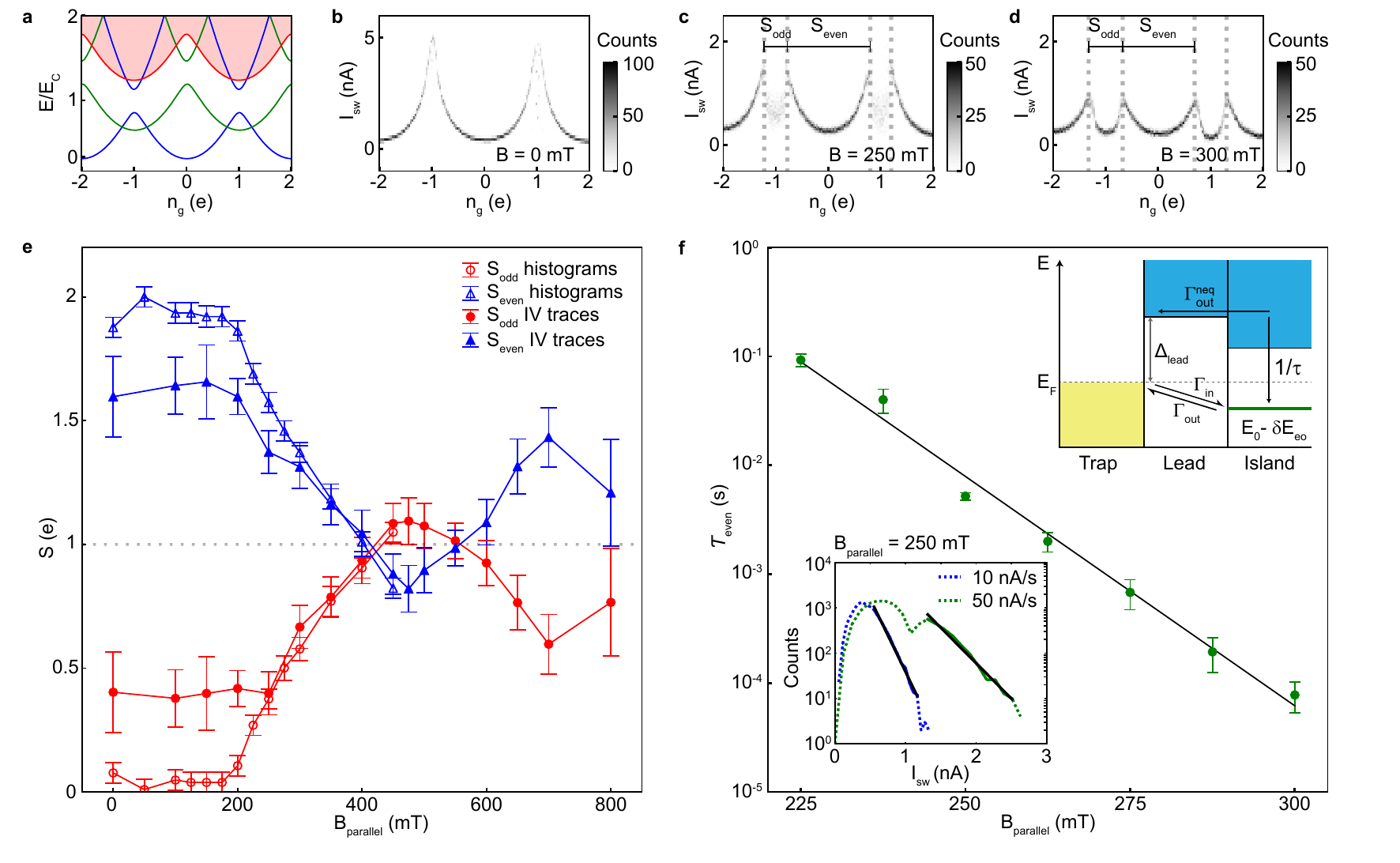}
\caption{\textbf{Parallel magnetic field dependence.} \textbf{(a)} The many body energy levels of the SCPT at finite magnetic field. The blue lines indicate the ground and first excited state of the even parity branch. The odd parity ground state is shown in red with the shaded red region emphasizing the quasiparticle continuum on the island. The green curves indicate the presence of a discrete subgap state on the island with energy $E_0(B)$ leading to an even-odd structure of the ground state when $E_0<\delta E_\text{eo}$ \textbf{(b-d)} Switching current histograms at 0 mT, 250 mT, and 300 mT showing the field evolution of the even-odd structure. \textbf{(e)} The even and odd spacings as a function of the parallel magnetic field obtained from both histograms and I-V traces. The observed crossing at 420 mT and subsequent oscillation is attributed to the presence of a subgap state that oscillates about zero energy as a function of magnetic field while the superconducting bulk on the island remains gapped. For the most part switching current histograms and I-V characteristics give the same spacings. At low fields the slow I-V measurements pick up rare poisoning events and thus do not recover full $2e$ periodicity \textbf{(f)} The even parity lifetime at $n_g=1$ as a function of the magnetic field. The solid line is a guide to the eye indicating an exponential dependence.  The lower inset presents a typical dataset used for the extraction of $\tau_\text{even}$. The upper inset shows a schematic representation of the energy needed to add a single quasiparticle to different parts of the device.}
\label{Fig4}
\end{center}
\end{figure*}

The other devices behave similarly as can be seen in Fig. S1 of the Supplementary Information. Five out of the six measured SCPTs show a 2$e$-period modulation robust over different gate settings. The remaining SCPT (device 5) exhibits an even-odd pattern, indicating that $\delta E_\text{eo} > \Delta$. Nevertheless, the robustness of the 2$e$-signal across gate settings and devices suggests a general lack of low-energy subgap states inside the islands at zero field, consistent with the hard gap observed in bias spectroscopy experiments which locally probe the density of states \cite{chang2015hard, deng2016majorana}.

\textbf{Temperature dependence and modeling.}  To gain insight into the relevant poisoning mechanisms of the SCPT, we measure the temperature dependence of the 2$e$-periodic switching current modulation at $R_N = \SI{14.8}{\kilo\ohm}$. As can be seen in Fig. \ref{Fig3}a, we observe that the 2$e$ periodicity persists up to $T \approx 189$ mK at which point the oscillations develop local maxima at even $n_g$ values and finally become fully $1e$-periodic for $T^* \approx 300$mK. This is consistent with an expected level spacing $\delta$ of the Al shell of a few mK when using the estimate for vanishing charge dispersion $k_B T^*=\Delta/\ln(\Delta/\delta)$ \cite{Matveev1994}. For comparison to the histograms, Fig \ref{Fig3}b shows $dV/dI$ data taken over the same temperature range. At elevated temperatures the $dV/dI$ characteristics show a similar behavior as the histograms including the onset of local maxima at even $n_g$. This can be explained by a self-averaging that takes place in the overdamped regime due to a succession of multiple switching and retrapping events. Indeed, we note that for $T>189$ mK, the $dV/dI$ traces show negligible hysteresis, indicating that the SCPT is in the overdamped regime. At low temperatures the junction enters the underdamped regime where a single phase slip can drive the junction normal, which leads to an increase of noise in the $dV/dI$ data at base temperature.

Our modeling of the $dV/dI$ data, outlined in Supplementary Information Section I, focuses on the overdamped regime. We identify two limiting cases, depending on the ratio of the parity switching times controlled by $1/\Gamma_\text{in}, 1/\Gamma_\text{out}$ and the response time of the SCPT given by the Josephson time constant $\tau_J = \hbar/2eI_cR_J$ \cite{kautz1990noise}, with $R_J$ the effective shunt resistance of the device and $I_c$ the critical current. For slow parity switches one expects a double peak structure in the $dV/dI$. In contrast we observe a parity-averaged single peak in the $dV/dI$ which shows that at high temperatures the SCPT is in the fast parity switching regime $\Gamma_\text{in}, \Gamma_\text{out} \gg 1/\tau_J$. At $T \approx 189$ mK where the SCPT transitions into the overdamped regime, $R_J \approx \SI{180}{\ohm}$ and $I_c \approx 3$ nA leading to $\tau_J \approx 1$ ns in our experiment.

Given the fast (un)poisoning at high temperature, we model the observed switching currents as the weighted sum of the switching current of the even and the odd parity states, with the relative probabilities governed by the free energy difference of the two states. Our model includes the charging energy of the island, Josephson coupling of the island to the leads, and the entropic factor associated with bringing a quasiparticle into the island; see Supplementary Information Section II for a more detailed discussion.  
We note that though the fast (un)poisoning is a necessary assumption to fit the data at high temperature $T > 189$ mK, at low temperatures the probability to find the system in the odd state becomes negligible, i.e. $p_\text{odd}/p_\text{even}\propto\exp(-(\Delta-\delta E_\text{eo})/k_B T)\to 0$ for $\Delta>\delta E_\text{eo}$. Thus, for low temperatures the system is essentially only in the even state which yields the $2e$-periodic histograms of Fig~\ref{Fig3}a (v).

The fitting gives approximate values of the $\Delta$, $E_J$, and $E_C$. These values have error bars of the order of half of their values due to the weak parameter dependence of the fitting function. The fitted value of the superconducting gap $\Delta \approx \SI{220}{\micro\electronvolt}$ is, within its error bar, consistent with the value obtained from the Coulomb diamonds in Fig. \ref{Fig2}a. Similarly, the fitted $E_J \approx \SI{43}{\micro\electronvolt}$ is consistent with the observed switching current. The fitted effective $E_C \approx \SI{160}{\micro\electronvolt}$, however, is smaller than $E_C^{0}$ extracted from the Coulomb diamond data in Fig. \ref{Fig2}a. This indicates that, in the regime of open tunnel barriers, $E_C$ is significantly renormalized by virtual quasiparticle tunneling processes relative to the geometric charging energy \cite{Ambegaokar82,Larkin83,Eckern84,Lutchyn07}. The set of consistent fit parameters, together with an excellent fit of the model to the observed switching current dependence on $n_g$, supports the validity of the model and the assumption of fast (un)poisoning at high temperatures. Similar fitting results for device 2 strengthen this conclusion, see Supplementary Information Fig. S4.

\textbf{Parallel magnetic field dependence.} Next, we study the effect of a parallel magnetic field on the switching current modulation. In particular, we tune the gates such that $R_N = \SI{12.9}{\kilo\ohm}$ and $I_{sw}$ shows a 2$e$-periodic modulation at zero field, as is shown in Fig. \ref{Fig4}b.  The 2$e$ periodicity implies that $\Delta>\delta E_\text{eo}$ and thus that the ground state is always even. As a magnetic field is applied along the nanowire axis, the spinful, odd-charge states are split by the Zeeman energy, thereby reducing the minimal single particle excitation energy $E_0$ in the island. Note that the effective g-factor of a subgap state residing partially in the InAs nanowire may be larger than that of the states in the Al shell \cite{vaitiekenas2017effective}. Therefore, Fig. \ref{Fig4}a shows a single state (in green) detached from the quasiparticle continuum appearing in a field. Interestingly, when the applied field is large enough so that $E_0 <\delta E_\text{eo}$, the parity of the ground state around $n_g=\pm1$ changes to odd. During the retrapping process of the switching current measurement the system tends to be reset to the ground state. Hence, the corresponding parity-flip shows up as a dip in the switching current modulation around odd $n_g$, causing an even-odd pattern. Figures \ref{Fig4}c and \ref{Fig4}d show examples of this even-odd structure in the switching current modulation measured at 250 mT and 300 mT, respectively. 

We investigate the field dependence of this even-odd pattern in more detail by defining the length in gate-charge over which the even (odd) state is stable as $S_\text{even}$ ($S_\text{odd}$). In Fig. \ref{Fig4}e these spacings are tracked as a function of the magnetic field using both switching current histograms and I-V measurements, see Fig. S5 and S6 of the Supplementary Information for the representative data. The even (odd) data points are obtained by averaging over 2 (3) successive spacings. Unlike earlier studies in Al-AlO$\mathrm{_x}$ SCPTs \cite{lafarge1993evenodd, tuominen1993field}, the even and odd spacings cross at 420 mT. After the first crossing, the spacings oscillate around 1$e$ with an increasing oscillation amplitude. The maximum oscillation amplitude at 700 mT sets a lower bound on the bulk superconducting gap at lower fields since the bulk gap can be assumed to decrease monotonically with field. Therefore, we conclude that the oscillating spacing indicates the zero energy crossing of a single subgap state that is detached from the continuum as is illustrated in Fig. \ref{Fig4}a.

Similar to Fig.~\ref{Fig3}, the histograms and I-V characteristics mostly coincide. For small fields below 200 mT, however, the histograms indicate an even ground state, while the slower I-V traces display an even-odd pattern, see Fig. \ref{Fig4}e. This discrepancy occurs because the slower I-V measurements are sensitive to rare trapping events of quasiparticles in the island \cite{lutchyn2007kinetics}. The latter occur since even in the absence of subgap states the island acts as a metastable trap with energy $\delta E_\text{eo}$ below the gap of the superconducting lead around odd $n_g$. In rare cases the metastable state becomes occupied long enough by quasiparticles to cause switching to the resistive state.

In addition, we measure the parity lifetime of the SCPT in a parallel field by performing slow histogram measurements while fixing the gate-charge at $n_g = 1$ so that the extracted lifetime corresponds to poisoning of the even state \cite{mannik2004effect, van2015one}. For representative histograms see the lower inset of Fig. \ref{Fig4}f and Fig. S7 of the Supplementary Information. At $n_g=1$, we expect the worst-case scenario for poisoning since the energy difference between the even and odd state is maximal (i.e. favoring the odd state). We observe that this lifetime decreases exponentially with field between 225 and 300 mT, see Fig. \ref{Fig4}f. We are limited to this intermediate field range because the lifetime is too large to obtain useful statistics at lower fields and too small to be captured by the bandwidth of the measurement electronics at larger fields. Still, by extrapolating the lifetime to 415 mT where $S_{\mathrm{even}} = S_{\mathrm{odd}} = 1e$, one can estimate the parity lifetime when the subgap state is at zero energy to be $\approx 1$ ns.

\section{Discussion}

We begin by noting that the growth of the even-odd spacing oscillation as a function of field seen in Fig. \ref{Fig4}e is reminiscent of one of the proposed signatures of overlapping Majorana zero modes \cite{sarma2012smoking}. However, this increasing oscillation amplitude was only observed in a narrow gate range in our device, as is illustrated in Fig. S8 of the Supplementary Information. This makes it difficult to map the amplitude of the first oscillation to a Majorana overlap, as was done in Ref. \citep{albrecht2016exponential}.  From our results we can only conclude that if this oscillation is indeed due to the presence of overlapping MZMs, the topological portion of the device parameter space is rather small.  Nevertheless, mapping the even-odd peak spacing in this manner could be used in future experiments to signal the transition to the topological regime in devices with superconducting leads such as the ones proposed in Refs. \cite{aasen2016milestones, hell2016time,litinski2017combining}. This could be an attractive alternative to gap-edge spectroscopy \cite{PhysRevB.95.020501,San-Jose2013,Nadj-Perge2014,Ruby15} as a signature of the topological regime in these all-superconducting systems.  

We also note that the splitting of the 2$e$-signal into an (oscillating) even-odd signal is not always observed. Measurements performed on device 4, which has a $\SI{3}{\micro\meter}$-long island, show a sharp transition of the 2$e$-signal to the 1$e$-signal at a parallel field of 100 mT, similar to the behavior observed while increasing the temperature in device 1, see Fig. S9 in the Supplementary Information. This field evolution of the $I_{sw}$ modulation indicates that the SCPT is in the fast (un)poisoning limit with $\Gamma_\text{in}/\Gamma_\text{out} \approx 1$, possibly caused by a field-induced softening of the superconducting gap in the island and/or leads.

To understand the exponential decrease of the even state lifetime with field seen in Fig~\ref{Fig4}f, we model the system as an island connected to a gapped superconducting lead in contact with a normal metal quasiparticle trap as is shown in the upper inset of Fig. \ref{Fig4}f. In the field range where we measure the lifetime, the observed even-odd pattern indicates that the energy difference between the odd and even state at $n_g = 1$ is always negative, as also depicted in Fig. \ref{Fig4}a and the inset of Fig. \ref{Fig4}f. Therefore, at $n_g=1$ poisoning is only prevented by the quasiparticle filtering effect of the superconducting gap in the leads. Quasiparticles can cross this gap in two ways: by thermal excitation to the gap edge, which is suppressed by a factor $\exp\left(-\Delta_\text{lead}/k_BT\right)$, or by tunneling through the gap, which is suppressed by a factor $\exp\left(-L/\xi\right)$.  Quantitative estimates of the relative strength of the tunneling and thermal activation contributions require a microscopic knowledge of the device.  However, both processes lead to an exponential dependence of the lifetime with field since for thermal activation $\Delta_\text{lead}(B) = \Delta_\text{lead}(0) - \frac{1}{2}g\mu_BB$, and similarly for tunneling, $\xi^{-1}\propto \Delta_\text{lead} \,(\sqrt{\Delta_\text{lead}})$ for clean (dirty) superconductors \cite{tinkham1996introduction}. In either case, the filtering effect should be enhanced by increasing the length of the superconducting leads as well as by increasing $\Delta_\text{lead}$. Since recent studies indicate that the size of the proximitized gap in semiconducting nanowires is gate-tunable \cite{vaitiekenas2017effective}, we suggest enhancing this filtering effect by locally gating the leads of the SCPT.

Next, for a Majorana-based qubit one is primarily concerned with poisoning events which change the state of the qubit - namely, poisoning of the MZMs.  If direct tunneling from the quasiparticle trap is the dominant poisoning mechanism, the subgap state is expected to be directly poisoned since it is the lowest energy state on the island. In this case, the measured $\tau_\text{even}$ in Fig. \ref{Fig4}f directly gives the bound state lifetime since the quasiparticle residence time in the subgap state is likely to be longer than the relevant switching timescale of the junction - $\tau_J$ in the case of an overdamped junction and $2\pi/\omega_p$ where $\omega_p$ is the plasma frequency in the case of an underdamped junction. In the opposite case of thermally activated poisoning, quasiparticles are first transfered elastically from the superconducting lead to the continuum in the island before relaxing to the subgap state within a time $\tau$ \cite{lutchyn2007kinetics}. In this case, quasiparticles can escape from the island before they are detected if $\Gamma_\text{out}^{\text{neq}}$ is faster than the SCPT response time. Note, however, that as long as quasiparticles can be detected faster than $\tau$, most of the poisoning events of the subgap states will be detected. The time  $\tau_\text{even}$ therefore again represents the parity lifetime of the subgap states while the overall parity of island might fluctuate faster. Our previous estimate of $\tau_J \approx 1$ ns sets a lower bound on our poisoning detection bandwidth since the junction would switch even faster to the resistive state in the underdamped case which we observe at low temperature.  Given that typical resonators in time-domain RF measurements have bandwidths of no more than a few 10's of MHz \cite{naaman2006time,shaw2008kinetics,court2008quantitative}, switching current measurements are a promising alternative before Majorana poisoning times can be measured more directly via the coherence of MZM-based qubits.

Finally, with the design of future MZM-based qubits in mind it is worth comparing our results with those obtained with NbTiN islands \cite{van2015one}.  Our observed gate-charge modulation of the switching current shows a robust 2e-periodic signal for a wide range of gate settings which indicates that there are no low-energy subgap states inside the SCPTs at zero magnetic field. To emphasize this point, even though the island is flooded with quasiparticles after each measurement, the parity of the system consistently resets to the same value. This is in stark contrast to the case of NbTiN islands, where subgap states result in a 1$e$-periodic, bimodal switching current distribution.  In that case, despite the large superconducting gap, the island parity is effectively randomized after each measurement when the island retraps after being flooded with quasiparticles.  Thus, with an aim of keeping quasiparticles as far from fragile MZMs as possible, we conclude that Al is a better candidate than NbTiN for future qubit devices.  Additionally, our observed exponential decrease of the quasiparticle lifetime with increasing magnetic field highlights the importance of proper engineering of the superconducting gap even when using Al films with a hard superconducting gap. We suggest incorporating local gating and intentional quasiparticle traps to minimize the presence of quasiparticles in the leads for qubit implementations requiring quenching of the island charge dispersion.

\section*{Methods}

\noindent
\textbf{Nanowire growth and device fabrication.} InAs nanowires are grown by molecular beam epitaxy via gold-catalyzed vapor-liquid-solid growth. A thin aluminium shell is then evaporated on two facets of the nanowire before breaking vacuum in order to form a pristine semiconductor-aluminium interface. The InAs-Al nanowires are deposited deterministically on a Si++ substrate covered with 285 nm thick thermal SiO$_\mathrm{2}$ using a micro-manipulator. The Josephson junctions together with the island are created by etching the Al shell from the nanowire in two 70 nm wide windows using a 12 s Transene D dip at $\SI{48}{\celsius}$. The wire is contacted $\SI{1}{\micro\meter}$ away from the junctions using an argon plasma etch at 100 W for 2 min and 45 s to remove the native oxide, followed by the \textit{in situ} deposition of NbTi/NbTiN (5 nm/70 nm) and the \textit{ex situ} deposition of Ti/Au (5 nm/35 nm). Finally, Ti/Au (10 nm/120 nm) side gates are deposited using evaporation. For all lithography steps electron beam lithography was used to pattern the resist (PMMA 950k A4 spin coated at 4000 rpm). For device 6, Ti/Au (5 nm/10 nm) local back gates and a 30 nm thick SiN$_x$ dielectric layer were deposited prior to the nanowire deposition.  

\noindent
\textbf{Switching current histograms.} The switching current histograms were measured using a Rigol DG4062 arbitrary waveform generator to supply a sawtooth waveform to an optically isolated current source which results in a time-dependent current bias of the device characterized by a constant current ramp rate $dI/dt$. The voltage across the SCPT is measured in a four-terminal configuration using a voltage amplifier that is optically isolated from the commercial electronics. A typical current bias waveform together with the resulting voltage are schematically depicted in Fig. \ref{Fig2}c. When the measured voltage crosses the preset voltage threshold $V_{th}$, the corresponding bias current is recorded using a custom sample-and-hold circuit and a Keithley 2000 digital multimeter. This reference voltage is tuned inside the voltage step that separates the supercurrent branch from the quasiparticle current branch so that the recorded current measures the switching current. This measurement is repeated $N$ times to acquire the switching current histogram. The readout lines consist of Cu/Ni twisted pair cables, and are filtered using three stages of filtering: a $\mathrm{\pi}$ filter with a cutoff frequency of ~100 MHz at room temperature, high frequency copper-powder filters at base temperature, and two-pole $RC$ filters with a cutoff frequency of 50 kHz also at base temperature.

\bibliographystyle{naturemag_noURL}
\bibliography{bib_2eSCPT}

\section*{Acknowledgements}
The authors thank D. J. van Woerkom for useful discussions and R. N. Schouten and O. Benningshof for technical assistance. This work has been supported by the Netherlands Organization for Scientific Research (NWO), Microsoft Corporation Station Q, the Danish National Research Foundation, and a Synergy Grant of the European Research Council.

\section*{Author contributions}
J.v.V., A.P., A.G., and J.D.W. designed the experiment.  J.N. and P.K. grew the nanowires.  J.v.V., A.P., and J.D.W. fabricated the devices and performed the measurements.  T.K., D.P., R.L., and J.v.V. performed the theoretical modeling.  All authors contributed to analyzing the data and writing the paper.

\end{document}